\documentclass[12pt]{article}
\usepackage{graphics}

\newcommand{\be}{\begin{equation}}
\newcommand{\ee}{\end{equation}}
\newcommand{\bea}{\begin{eqnarray}}
\newcommand{\eea}{\end{eqnarray}}

\begin{document}

\begin{center}
{\Large\bf Wannier-functions characterization\\of floating bonds in
a-Si}

\vspace{0.5cm}
{ M. Fornari,$^{\rm (1,2)}$ 
N. Marzari,$^{\rm (1)}$ M. Peressi$^{\rm (2)}$ and
A. Baldereschi$^{\rm (2,3)}$} \\ 
\vspace{0.6cm}
{\it $^{\rm (1)}$ Naval Research Laboratory, Washington, DC 20375 and CSI George Mason University, Fairfax, VA 22030}\\
{\it $^{\rm (2)}$ INFM and Dipartimento di Fisica Teorica - Universit\`a di
Trieste, Strada Costiera 11, I-34014 Trieste, Italy}\\
{\it $^{\rm (3)}$ Institut de Physique Appliqu\'ee, Ecole Polytechnique
F\'ed\'erale de Lausanne \\ PHB-Ecublens, CH-1015 Lausanne,
Switzerland}
\end{center}

\begin{abstract}
We investigate the electronic structure of 
over-coordinated defects in amorphous silicon via density-functional
total-energy calculations, with the aim of understanding the relationship
between topological and electronic properties on a microscopic scale.
Maximally-localized Wannier functions 
are computed in order to
characterize the bonding and the electronic properties of these defects.
The  five-fold coordination defects
give rise to delocalized states extending over several
nearest neighbors, and therefore to very 
polarizable bonds and anomalously high Born effective charges 
for the defective atoms.

\end{abstract}

\section{Introduction}

Coordination is a well-defined concept in crystalline systems, 
where it is closely related to the concept of bonding.
For instance, silicon atoms are 
tetrahedrally coordinated, and  exhibit covalent bonds with their
four nearest-neighbors (NNs) in the ordinary crystalline phase (c-Si).
Whenever symmetry is broken, as in the case of amorphous silicon (a-Si) 
studied here, the link between coordination and bonding becomes less 
obvious. Even if atomic positions are intimately connected with the
electronic 
ground state, we are usually unable to extract detailed informations 
about bonding from a purely geometrical investigation. Standard 
analytical tools based on coordination-number analysis are thus
insufficient to characterize bonding properties, since they are
insensitive to the details of the electronic charge distribution.
Whenever strong topological disorder is present,
as in the case of atoms with three or five nearest neighbors 
(T$_3$ or T$_5$ defects), the ionic potential and the charge 
distribution can be significantly different from the crystalline case,
and great care should be used to identify the nature of 
bonding.

In this work we use density-functional theory in the local density
approximation to calculate the electronic properties of over-coordinated 
defects in a-Si. Density-functional theory provides 
an accurate and parameter-free description of the electronic
ground-state, and is equally capable of dealing with four-fold coordinated
atoms and with more complex topologies.
The characterization of defects is then based on the decomposition
of the electronic ground state into localized orbitals, 
using the technique of maximally-localized Wannier 
functions (MLWF) \cite{MV}. 
In this approach, the extended Bloch orbitals are transformed via 
unitary transformations into a representation where they are maximally
localized. The role and importance of such localized Wannier functions in
the study of disordered systems (amorphous silicon in particular) 
has been advocated  by Silvestrelli and coworkers \cite{SP},
who have shown that the coordination analysis 
is often insensitive to the electronic charge distribution and that
similarly coordinated atoms can be surrounded by rather  
different bonding environments. Our present results support 
these findings.

The conjecture that over-coordination could play an important role in the
formation of mid-gap electronic levels in amorphous silicon
has been recently validated by some of us \cite{EPL},
using accurate density functional calculations.
In that work we argued that T$_5$ defects can 
be responsible, as much as T$_3$ ones, for states close to the
Fermi level. 
We also showed that in the case of T$_3$ defects the mid-gap electronic 
state originates from the dangling bond and  is well localized on the
T$_3$ defect itself. Conversely, in T$_5$ defects the  midgap      
electronic state is delocalized over several
NNs of the T$_5$ site, in agreement with the findings of tight-binding
calculations reported in Ref. \cite{T5}.
In this work we pursue further the study of the extended nature 
of T$_5$ defects. Our goal is to understand how many atoms can be
actively involved in a defect, since this number influences the 
shape of the ``super-hyperfine'' structure of the D center 
in the electron-spin resonance (EPR) spectrum\cite{S}. In turn, this should
discriminate dangling bonds from floating ones. It should be noted
that both T$_3$ and T$_5$ centers generate a D signal; it is the
contributions of the secondary ions involved in the uncompensated
spin distribution (the super-hyperfine structure) that should be
very different in the case of a dangling or a floating bond. 

The plan of the work is as follows:
in Sec. 2 we describe our {\sl ab-initio}
calculations and we review the method used to compute the
MLWFs. We also comment on the link with
the dynamical Born effective charges $Z$. In Sec. 3 we 
discuss the results for our a-Si sample,
in comparison with the results obtained using the 
``electron-localization function''
(ELF) \cite{PhM} and the ``atomic-projected charge''
(APC) analysis \cite{EPL}. Sec. 4 is devoted to the conclusions.

\begin{figure}
\begin{center} 
\begin{minipage}{14cm}
 \hspace{-1cm}
 \rotatebox{0}{\scalebox{.75}[.75]{\includegraphics{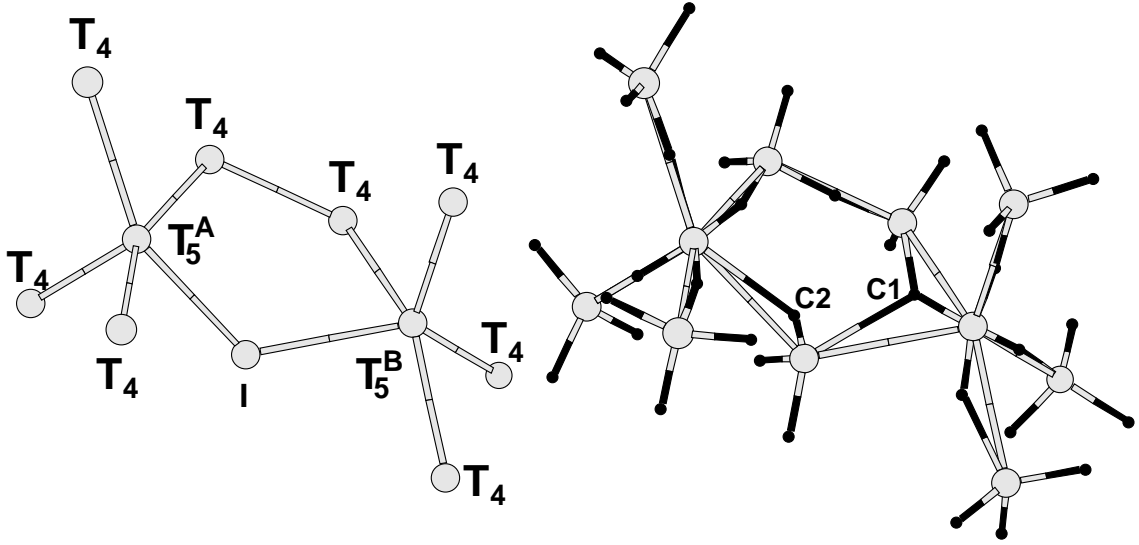}}} 
\end{minipage}  
  \caption{\small Left panel: structure of the 
  ``defect group'' embedded in our a-Si sample. Atoms are labelled
  T$_5$ (T$_4$) when they are five-fold (four-fold) coordinated.  
 Atom I (interstitial) is the only four-fold
  atom that is a NN of both T$_5$'s. Right panel: the ``defect
  group'' is shown together with the centers of the associated Wannier 
 functions.  C1 and C2 are the  centers with the lowest symmetry, and
  exhibit 
 the strongest deviations  
 from the geometrical mid-bond positions.}\label{FIG1}
\end{center}
\vspace{-0.8cm}
\end{figure}

\section{Computational tools}

The calculations performed here are based
on density-functional theory in the local-density approximation,
using a cubic super-cell containing 64 Si atoms \cite{BCP}, a 
plane wave basis set, and accurate
norm-conserving pseudopotentials. 
The electronic ground-state is obtained using
an all-bands conjugate gradient minimization \cite{eDFT}.
The sampling in reciprocal space is performed with 8 {\bf k} 
points in the full Brillouin zone. 
 
The analysis of the Kohn-Sham orbitals and their charge density
has been previously performed 
using the APC and the ELF analysis \cite{EPL,PhM}. 
APC is a measure of the contribution of the different atoms 
to the charge density, and it is obtained projecting the Bloch orbitals
$\Psi_{n{\bf k}}$ on localized atomic functions \cite{EPL}.
ELF, defined as in Ref. \cite{ELF}, 
is a measure of the conditional probability of 
having one electron close to another one with the same spin. It approaches
its upper limit (1.0) when the electron density resembles a covalent bond
or in the presence of unpaired electrons. 

Both these approaches provide
useful results in the characterization of
the electronic distribution; still, there are several
limitations that will be mentioned in the next section. 
The analysis based on localized Wannier functions  overcomes 
these limitations; as a byproduct, it also provides informations 
on the dielectric properties of the system.

Wannier functions are an alternative
representation of the electronic ground state,
that classifies states using a
spatial coordinate {\bf R} and a band index $n$. Wannier functions
are not eigenstates of the Kohn-Sham
Hamiltonian, but are related to them via the 
unitary transformation
\be
\label{eq:wannier}
|{\bf R}n \rangle = {V \over {(2\pi)^3}} \int_{BZ} |\Psi_{n{\bf
 k}} \rangle e^{-i{\bf R}\cdot{\bf k}} d{\bf k}  .
\ee
In the electronic structure problem there are 
degrees of freedom that do not affect the self-consistent ground state,
but  determine the shape of the WFs obtained 
from Eq. (\ref{eq:wannier}). 
They consist of arbitrary unitary rotations $U_{mn}({\bf k})$ that mix
together at any given ${\bf k}$ in the Brillouin zone
fully occupied Bloch orbitals 
(in the case of a single band, these unitary matrices reduce
to a phase factor $\phi({\bf k})$).
It is of paramount importance, in order
to provide a meaningful real-space representation, to choose these
arbitrary matrices $U_{mn}({\bf k})$ so that the resulting WFs 
are well localized.
The approach used here follows the lines of Ref. \cite{MV},
where the unitary rotations are refined until the resulting 
Wannier functions
are maximally-localized, i.e. they have  minimum spread 
\be
\Omega = \sum_n [ \langle {\bf 0}n| r^2 |{\bf 0}n \rangle -
\langle {\bf 0}n| r |{\bf 0}n \rangle^2]\,.
\ee

Incidentally, the sum of the Wannier function centers (WFCs)
${\bf r}^c_n=\langle {\bf 0}n| {\bf r} |{\bf 0}n \rangle$
is directly related to 
the macroscopic polarization of the sample \cite{VKS},
and this makes the Wannier function analysis attractive to
study dielectric properties.  In particular, the change in
polarization induced by the
displacement $\Delta{\mathbf \tau}_N$ of an atom N is directly
related to the electronic component of its Born dynamical-charge
tensor $Z_N$ by 
\be
(Z_N)_{i,j}= -2 {{\sum_n \Delta{\bf r}^c_n
\cdot {\bf e}_i} \over{ \Delta{\mathbf \tau}_N \cdot {\bf e}_j}}  
\ee
where $n$ runs over all the occupied  WFs in the unit cell
and $\Delta{\bf r}^c_n$ are the displacements of the
WFCs.
 
\begin{figure}
\begin{center} 
\begin{minipage}{8cm}
 \vspace{-1.cm} 
 \hspace{-4.5cm}
 \rotatebox{-90}{\scalebox{.7}[.7]{\includegraphics{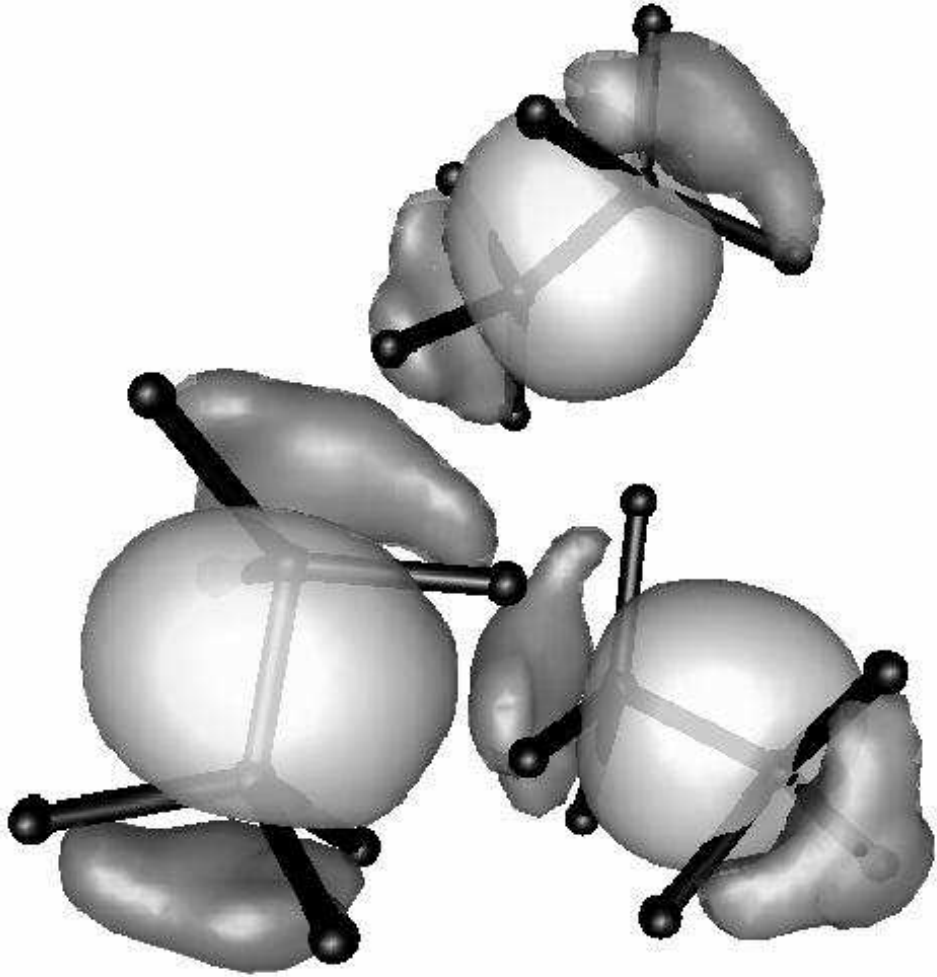}}} 
\end{minipage}  
   \caption{\small Wannier functions centered near crystal-like 
  coordinated atoms in the a-Si supercell. In each case a covalent
  bond (light gray) exists between two four-fold coordinated
  atoms, and the WFC is  close to the geometrical bond center. 
  For each atom, clover back-bonds (dark gray) point towards the
  remaining three neighbors.}\label{FIG2}   
\end{center}
\vspace{-.8cm}
\end{figure}

\section{Results on a-Si}

The presence of mid-gap states in a-Si is usually ascribed to the
existence of T$_3$ defects in the system. Such conclusion relies 
on EPR experiments, which measure the 
unpaired spin oscillations of the electrons in the
dangling bonds. Nevertheless, certain features of the EPR signals 
(the ``super-hyperfine structure'') derive also
from the ionic magnetic moment and the quadrupolar term of
the secondary ions in the defect \cite{S}. 
Thus, the EPR signal is indirectly sensitive to the presence of
complex geometries.
$T_5$ defects are expected to give rise to more delocalized states 
involving several atoms \cite{EPL,T5}, and could thus provide a
distinctive shape to the EPR signal. Prompted by these
considerations, we investigate in this work the electronic properties of
over-coordinated defects in amorphous silicon.

Our samples have been obtained by annealing
configurations spanned during a first-principles 
molecular dynamics simulation \cite{BCP}. In particular, we have selected
for this study 
a sample presenting two $T_5$ coordination defects.
Although this corresponds to a density of defects
higher than that of experiments, it still results in a useful model to 
investigate the effects of local strain and topological disorder.
The defects were first identified  using a geometrical analysis:
an atom is counted as a neighbor if it resides inside 
a sphere of radius $R_c$ = 2.6 $\rm \AA$ (to compare with the
theoretical NN distance in c-Si $d_{NN}$ = 2.357 $\AA$).  
This radius is
chosen as the first minimum in the correlation function $g({\bf r})$, 
since the peaks correspond, intuitively, to the succession of shells 
surrounding the reference atom. 
The two $T_5$ defects (labelled T$_{5}^A$ and T$_{5}^B$
in Fig.\,\ref{FIG1}) belong 
to a ``defect group'' (Fig.\,\ref{FIG1}, left panel), 
mostly composed of  bonds which are longer and weaker than four-fold ones. 
Such environment seems favorable to induce delocalization of
the wavefunction over several neighbors; our previous APC analysis
(see Ref.\,\cite{EPL}) supported this conjecture, as did a similar
investigation (Ref.\,\cite{PhM}) based on the
electronic localization function. 
Both approaches however suffer from inherent limitations: 
the projection technique used for the APC does not provide 
meaningful informations on the
mid-bond region; on the other side, ELF (once it has been been used to
unambiguously distinguish between a $T_3$ and a $T_5$ site) is not able   
to improve the qualitative description of the defect that can be obtained 
from the direct inspection of the charge density.

To improve our understanding of the electronic structure of the
defect group, we calculated the MLWFs for this
configuration. In crystalline Si each WF is centered in the middle of the bond
and oriented along the bond \cite{MV}; a covalent ``barrel''
of charge is shared between the two bonded atoms, 
while some back-bonding charge is distributed
between each of these atoms and its remaining three NNs.
We expect that in a slightly distorted environment, where short range
order is conserved, the shape of the WFs would not drastically change.
This is clearly confirmed by our calculations (see Fig. \ref{FIG2}). 
Small deviations from the crystalline case are due to the bending 
and stretching of bonds, but no doubt exists on  the nature of
the bond, that resembles closely to the bond of crystalline silicon.
Such ``regular'' bonds are clearly related to their crystalline
counterparts by having a very peaked distribution in 
their spreads, with a maximum around 2.30 $\rm \AA^2$
(vs. 2.04 $\rm \AA^2$ for crystalline silicon, using the same 8 {\bf
k}-point sampling). 

\begin{figure}
\begin{center} 
\begin{minipage}{14cm}
 \vspace{-1cm} 
 \hspace{-1cm}
 \rotatebox{-90}{\scalebox{.38}[.38]{\includegraphics{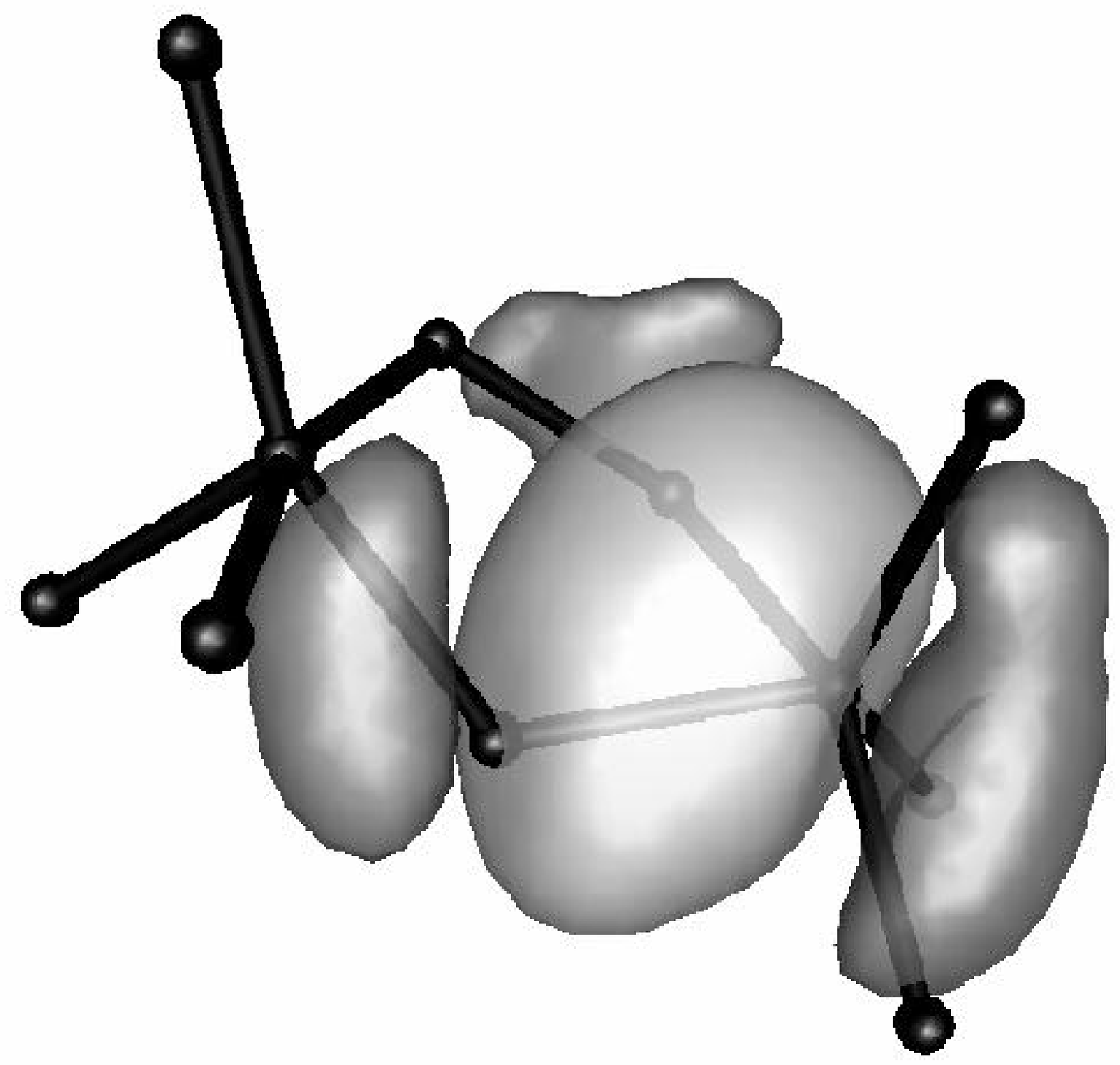}}} 
 \hspace{-1.cm}
 \rotatebox{-90}{\scalebox{.38}[.38]{\includegraphics{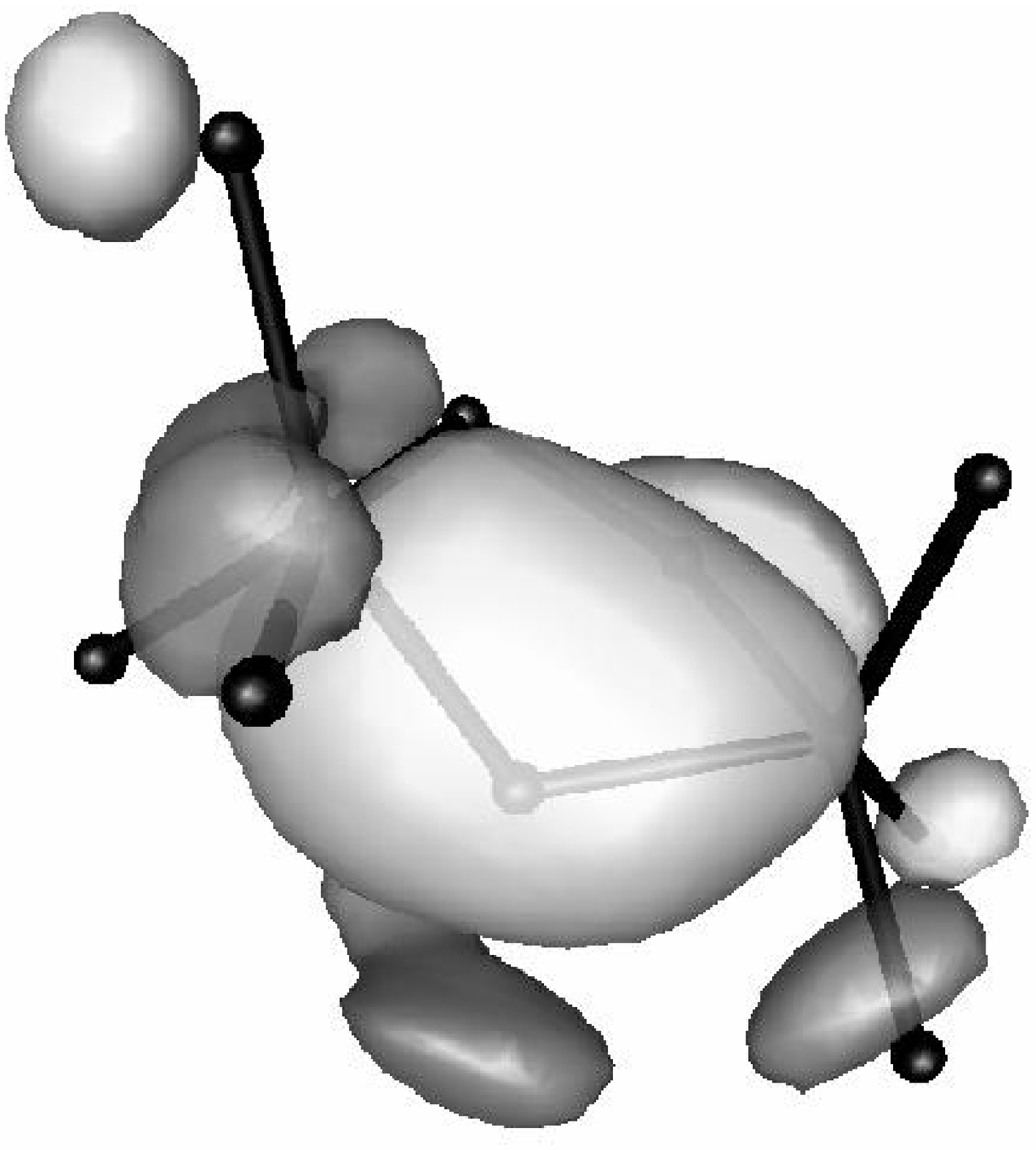}}}
\end{minipage}  
  \caption{\small Wannier functions corresponding to two bonds
  included in the ``defect group'' of Fig.\,\ref{FIG1}; we show
  the WF centered at C1 (left panel) and C2 (right panel). C1 occurs
  between two bonds, and the corresponding WF is delocalized
  over the $T_5$ site and two of its nearest neighbors.
}\label{FIG3}   
\end{center}
\vspace{-0.8cm}
\end{figure}

We started our analysis focusing on the position
of the WFC, as suggested by Silvestrelli et al. \cite{SP}. 
In that work it was observed that the centers of charge for
anomalous WFs tend to be closer to one specific atom than those of
regular WFs, which are instead equally distant from two bonded atoms.
In Fig.\,\ref{FIG1} we show the WFCs belonging to
our ``defect group''; it can be seen that around the 
interstitial atom I the Wannier centers 
depart strongly from their ideal mid-bond positions. 
Similarly, the shape 
of the corresponding Wannier functions is 
clearly anomalous, with a delocalization extending over
more than two atoms and with a much larger spread than average,
of the order of 4 to 8 $\rm \AA^2$.
This confirms our conjecture of delocalized orbitals connecting different 
atoms inside the ``defect group''; an inspection of
the shape of the WF clearly confirms the
delocalization of the electronic states. In Fig.\,\ref{FIG3} the left
panel shows the WF with the center close to the T$^B_5$ defect
(C1); its shape suggest a bond shared between two NN.
In the right panel of Fig.\,\ref{FIG3}
the delocalized orbital centered at C2 is shown. The shape of this
orbital is also influenced by the additional 
interaction between the two T$_5$ defects.
 
The anomalous shape and spread of some WFs present in our
sample suggest that unusual polarization properties could be present, and
could be related to topological disorder. For this reason, we
investigated the dynamical charge tensor $Z$, 
to quantify the local anisotropy and deviation from the
crystalline order. Our preliminary results suggest that disorder plays 
a considerable effect on the dynamical charges of a-Si.
 
Following Ref. \cite{PC},
we decompose the $Z$ tensor into an isotropic contribution
(corresponding to the $l=0$
spatial rotation representation), a $l=1$ antisymmetric one , and 
a $l=2$ traceless symmetric contribution.
In crystalline Si the effective charges are zero, and 
the electronic contribution (-4 times the identity matrix)
cancels exactly the ionic contribution (+4 times the identity matrix). 
This is not the case in amorphous silicon, and we find that atoms belonging to
topological defects exhibit a very different behavior.
To this purpose it is instructive to inspect
Table\,\ref{TAB1}, where the electronic effective charge tensors
of some selected atoms are given, together with their
decompositions.
Even for regular T$_4$ atoms (T$_4^A$ and T$_4^B$), 
the effective charges can show strong anomalies, although this could
be an artifact of the high density of defects in the sample.
The anisotropy around atom T$^B_5$ on the other hand is clearly shown in
Fig.\,\ref{FIG3} (right panel): $Z$ reflects the directionality in the
polarization of this floating bond. Even if
our results are still preliminary, they clearly show that
the deviations of the effective
charges from the crystalline value are noteworthy, and much larger than
expected.

\begin{center}
\hspace{-2cm}
\begin{table}%%%%%%%%%%% tab 1
\begin{tabular}{crrrrrrrrrr}
\hline
          
&Z$^{I}_{ii}$&Z$^{A}_{12}$&Z$^{A}_{13}$&Z$^{A}_{23}$&Z$^{S}_{11}$&Z$^{S}_{12}$&Z$^{S}_{13}$&Z$^{S}_{22}$&Z$^{S}_{23}$&Z$^{S}_{33}$\\   
T$_4^{A}$ &-3.860&-1.118&0.559&2.193&-1.264&-1.646&-0.368&-0.493&-0.330&1.758\\ 
T$_4^{B}$ &-5.732&-1.734&0.213&1.882&0.564&-1.903&-0.271&-2.582&-2.385& -2.018\\
I         &-3.850&1.089&-0.706&0.418&-1.403&0.765&0.493&-0.018&0.254& 1.421\\
T$_5^{B}$ &-5.544&-1.831&0.646&2.261&3.735&-1.059&-0.374&4.120&-2.647&0.385\\ 
\hline
\end{tabular}
\caption[Tab. 1]{Born effective charges (electronic)
for some selected atoms in the a-Si super-cell. The effective
charge tensor has been 
decomposed into an $l=0$ isotropic part (Z$^{I}_{ii}$=Tr(Z)/3),
a $l=1$ antisymmetric part (Z$^A$=(Z$-$Z$^T$)/2), and
a $l=2$ traceless symmetric part (Z$^S$=(Z+Z$^T$)/2$-$Z$^{I}$).
The superscript $T$ indicates
transpose; atoms are labeled as in Fig. 1.}
\label{TAB1}
\end{table}
\end{center}
 
\section{Conclusions}

We have presented our results on the microscopic features of floating 
bonds in a-Si, which have been obtained with a maximally-localized 
Wannier functions approach. 
We confirm the conjecture that T$_5$ defects are accompanied by
well-defined delocalized states. Such  states can be 
accurately characterized in terms of Wannier functions;
a quantitative  measure of delocalization is then provided by the 
the corresponding spreads.
The delocalized states correspond to anomalous covalent bonds 
expanding over  more than two atoms.
The dielectric properties are readily available as a byproduct of the
Wannier analysis; we find strongly anisotropic effective charges that are
significantly different from zero.

The authors wish to acknowledge INFM (Istituto Nazionale di Fisica
della Materia) for the ``Iniziativa Trasversale di Calcolo Parallelo'',
and David Singh and the Naval Research Laboratory, where 
part of this work was performed with the support of ONR.  We are also
grateful to S. de Gironcoli for insightful comments.

\end{document}